
\magnification=1200
\hsize=31pc
\vsize=55 truepc
\hfuzz=2pt
\vfuzz=4pt
\pretolerance=5000
\tolerance=5000
\parskip=0pt plus 1pt
\parindent=16pt
\font\fourteenrm=cmr10 scaled \magstep2
\font\fourteeni=cmmi10 scaled \magstep2
\font\fourteenbf=cmbx10 scaled \magstep2
\font\fourteenit=cmti10 scaled \magstep2
\font\fourteensy=cmsy10 scaled \magstep2
\font\large=cmbx10 scaled \magstep1

\font\eightrm=cmr8
\font\eighti=cmmi8
\font\eightbf=cmbx8
\font\eightit=cmti8

\font\eightsy=cmsy8
\font\sixrm=cmr6
\font\sixi=cmmi6
\font\sixsy=cmsy6

\def\tenpoint{\def\rm{\fam0\tenrm}%
  \textfont0=\tenrm \scriptfont0=\sevenrm
                      \scriptscriptfont0=\fiverm
  \textfont1=\teni  \scriptfont1=\seveni
                      \scriptscriptfont1=\fivei
  \textfont2=\tensy \scriptfont2=\sevensy
                      \scriptscriptfont2=\fivesy
  \textfont3=\tenex   \scriptfont3=\tenex
                      \scriptscriptfont3=\tenex
  \textfont\itfam=\tenit  \def\it{\fam\itfam\tenit}%
  \textfont\slfam=\tensl  \def\sl{\fam\slfam\tensl}%
  \textfont\bffam=\tenbf  \scriptfont\bffam=\sevenbf
                            \scriptscriptfont\bffam=\fivebf
                            \def\bf{\fam\bffam\tenbf}%
  \normalbaselineskip=20 truept
  \setbox\strutbox=\hbox{\vrule height14pt depth6pt
width0pt}%
  \let\sc=\eightrm \normalbaselines\rm}
\def\eightpoint{\def\rm{\fam0\eightrm}%
  \textfont0=\eightrm \scriptfont0=\sixrm
                      \scriptscriptfont0=\fiverm
  \textfont1=\eighti  \scriptfont1=\sixi
                      \scriptscriptfont1=\fivei
  \textfont2=\eightsy \scriptfont2=\sixsy
                      \scriptscriptfont2=\fivesy
  \textfont3=\tenex   \scriptfont3=\tenex
                      \scriptscriptfont3=\tenex
  \textfont\itfam=\eightit  \def\it{\fam\itfam\eightit}%
  \textfont\bffam=\eightbf  \def\bf{\fam\bffam\eightbf}%
  \normalbaselineskip=16 truept
  \setbox\strutbox=\hbox{\vrule height11pt depth5pt width0pt}}
\def\fourteenpoint{\def\rm{\fam0\fourteenrm}%
  \textfont0=\fourteenrm \scriptfont0=\tenrm
                      \scriptscriptfont0=\eightrm
  \textfont1=\fourteeni  \scriptfont1=\teni
                      \scriptscriptfont1=\eighti
  \textfont2=\fourteensy \scriptfont2=\tensy
                      \scriptscriptfont2=\eightsy
  \textfont3=\tenex   \scriptfont3=\tenex
                      \scriptscriptfont3=\tenex
  \textfont\itfam=\fourteenit  \def\it{\fam\itfam\fourteenit}%
  \textfont\bffam=\fourteenbf  \scriptfont\bffam=\tenbf
                             \scriptscriptfont\bffam=\eightbf
                             \def\bf{\fam\bffam\fourteenbf}%
  \normalbaselineskip=24 truept
  \setbox\strutbox=\hbox{\vrule height17pt depth7pt width0pt}%
  \let\sc=\tenrm \normalbaselines\rm}

\def\today{\number\day\ \ifcase\month\or
  January\or February\or March\or April\or May\or June\or
  July\or August\or September\or October\or November\or
December\fi
  \space \number\year}

\newcount\secno      
\newcount\subno      
\newcount\subsubno   
\newcount\appno      
\newcount\tableno    
\newcount\figureno   

\normalbaselineskip=20 truept
\baselineskip=20 truept
\def\title#1
   {\vglue1truein
   {\baselineskip=24 truept
    \pretolerance=10000
    \raggedright
    \noindent \fourteenpoint\bf #1\par}
    \vskip1truein minus36pt}
\def\author#1
  {{\pretolerance=10000
    \raggedright
    \noindent {\large #1}\par}}
\def\address#1
   {\bigskip
    \noindent \rm #1\par}
\def\shorttitle#1
   {\vfill
    \noindent \rm Short title: {\sl #1}\par
    \medskip}

\def\pacs#1
   {\noindent \rm PACS number(s): #1\par
    \medskip}
\def\jnl#1
   {\noindent \rm Submitted to: {\sl #1}\par
    \medskip}
\def\date
   {\noindent Date: \today\par
    \medskip}
\def\beginabstract
   {\vfill\eject
    \noindent {\bf Abstract. }\rm}
\def\keyword#1
   {\bigskip
    \noindent {\bf Keyword abstract: }\rm#1}
\def\endabstract
   {\par
    \vfill\eject}

\def\entry#1#2#3
   {\noindent
    \hangindent=20pt
    \hangafter=1
    \hbox to20pt{#1 \hss}#2\hfill #3\par}
\def\subentry#1#2#3
   {\noindent
    \hangindent=40pt
    \hangafter=1
    \hskip20pt\hbox to20pt{#1 \hss}#2\hfill #3\par}
\def\section#1
   {\vskip0pt plus.1\vsize\penalty-250
    \vskip0pt plus-.1\vsize\vskip24pt plus12pt minus6pt
    \subno=0 \subsubno=0
    \global\advance\secno by 1
    \noindent {\bf \the\secno. #1\par}
    \bigskip
    \noindent}
\def\subsection#1
   {\vskip-\lastskip
    \vskip24pt plus12pt minus6pt
    \bigbreak
    \global\advance\subno by 1
    \subsubno=0
    \noindent {\sl \the\secno.\the\subno. #1\par}
    \nobreak
    \medskip
    \noindent}

\def\appendix#1
   {\vskip0pt plus.1\vsize\penalty-250
    \vskip0pt plus-.1\vsize\vskip24pt plus12pt minus6pt
    \subno=0
    \global\advance\appno by 1
    \noindent {\bf Appendix 
#1\par}
    \bigskip
    \noindent}

\def\ack
   {\vskip-\lastskip
    \vskip36pt plus12pt minus12pt
    \bigbreak
    \noindent{\bf Acknowledgments\par}
    \nobreak
    \bigskip
    \noindent}
\def\figures
   {\vfill\eject
    \noindent {\bf Figure captions\par}
    \bigskip}
\def\figcaption#1
   {\global\advance\figureno by 1
    \noindent {\bf Figure \the\figureno.} \rm#1\par
    \bigskip}
\def\references
     {\vfill\eject
     {\noindent \bf References\par}
      \parindent=0pt
      \bigskip}
\def\refjl#1#2#3#4
   {\hangindent=16pt
    \hangafter=1
    \rm #1
   {\frenchspacing\sl #2
    \bf #3}
    #4\par}
 \def\numrefjl#1#2#3#4#5
   {\parindent=40pt
    \hang
    \noindent
    \rm {\hbox to 30truept{\hss #1\quad}}#2
   {\frenchspacing\sl #3\/
    \bf #4}
    #5\par\parindent=16pt}
\def\refbk#1#2#3
   {\hangindent=16pt
    \hangafter=1
    \rm #1
   {\frenchspacing\sl #2}
    #3\par}
\def\numrefbk#1#2#3#4
   {\parindent=40pt
    \hang
    \noindent
    \rm {\hbox to 30truept{\hss #1\quad}}#2
   {\frenchspacing\sl #3\/}
    #4\par\parindent=16pt}

\catcode`\@=11
\def\vfootnote#1{\insert\footins\bgroup
    \interlinepenalty=\interfootnotelinepenalty
    \splittopskip=\ht\strutbox 
    \splitmaxdepth=\dp\strutbox \floatingpenalty=20000
    \leftskip=0pt \rightskip=0pt \spaceskip=0pt \xspaceskip=0pt
    \noindent\eightpoint\rm #1\ \ignorespaces\footstrut\futurelet\next\fo@t}
\def\ind{\hbox to 5pc{}}
\def\eq(#1){\hfill\llap{(#1)}}

\def\deqn#1{\displ@y\halign{\hbox to \displaywidth
    {$\@lign\displaystyle##\hfil$}\crcr #1\crcr}}
\def\indeqn#1{\displ@y\halign{\hbox to \displaywidth
    {$\ind\@lign\displaystyle##\hfil$}\crcr #1\crcr}}
\def\indalign#1{\displ@y \tabskip=0pt
  \halign to\displaywidth{\ind$\@lign\displaystyle{##}$\tabskip=0pt
    &$\@lign\displaystyle{{}##}$\hfill\tabskip=\centering
    &\llap{$\@lign##$}\tabskip=0pt\crcr
    #1\crcr}}
\catcode`\@=12


\message{cross referencing macros - BD 1991}

\catcode`@=11
\immediate\newread\xrffile\immediate\openin\xrffile=\jobname.xrf
\ifeof\xrffile
  \message{ no file \jobname.xrf - run again for correct forward references }
\else
  \immediate\closein\xrffile\input\jobname.xrf
\fi
\immediate\newwrite\xrffile\immediate\openout\xrffile=\jobname.xrf

\newcount\t@g
\def\order#1{\expandafter\expandafter\csname newcount\endcsname
   \csname t@g#1\endcsname\csname t@g#1\endcsname=0
   \expandafter\expandafter\csname newcount\endcsname
   \csname t@ghd#1\endcsname\csname t@ghd#1\endcsname=0

   \expandafter\def\csname #1\endcsname##1{\csname next#1\endcsname##1 }

   \expandafter\def\csname next#1\endcsname##1 %
     {\edef\t@g{\csname t@g#1\endcsname}\edef\t@@ghd{\csname t@ghd#1\endcsname}%
      \ifnum\t@@ghd=\t@ghd\else\global\t@@ghd=\number\t@ghd\global\t@g=0\fi%
     \ifunc@lled{@#1}{##1}\global\advance\t@g by 1%
       {\def\next{##1}\ifx\next\empty%
       \else\writenew{#1}{##1}\expandafter%
       \xdef\csname @#1num##1\endcsname{\t@ghead\number\t@g}\fi}%
       {\t@ghead\number\t@g}%
     \else
       \message{ Warning - duplicate #1 label >> ##1 << }
       \csname @#1num##1\endcsname%
     \fi}%

   \expandafter\def\csname ref#1\endcsname##1{%
     \expandafter\each@rg\csname #1eatspace\endcsname{##1}}

   \expandafter\def\csname #1eatspace\endcsname##1 ##2,%
    {\csname #1cite\endcsname##2 ##1 }

   \expandafter\def\csname #1cite\endcsname##1 ##2##3
     {\ifunc@lled{@#1}{##2##3}%
       {\expandafter\ifx\csname @@#1num##2##3\endcsname\relax%
         \message{ #1 label >>##2##3<< is undefined }>>##2##3<<%
       \else\csname @@#1num##2##3\endcsname##1\fi}%
     \else\csname @#1num##2##3\endcsname##1%
     \fi}}

\def\each@rg#1#2{{\let\thecsname=#1\expandafter\first@rg#2,\end,}}
\def\first@rg#1,{\callr@nge{#1}\apply@rg}
\def\apply@rg#1,{\ifx\end#1\let\next=\relax%
\else,\callr@nge{#1}\let\next=\apply@rg\fi\next}

\def\callr@nge#1{\calldor@nge#1-\end-}
\def\callr@ngeat#1\end-{#1}
\def\calldor@nge#1-#2-{\ifx\end#2\thecsname#1 ,%
  \else\thecsname#1 ,--\thecsname#2 ,\callr@ngeat\fi}

\def\writenew#1#2{\immediate\write\xrffile
     {\string\def\csname @@#1num#2 \endcsname{\t@ghead\number\t@g}}}

\def\ifunc@lled#1#2{\expandafter\ifx\csname #1num#2\endcsname\relax}

\def\t@ghead{}
\newcount\t@ghd\t@ghd=0
\def\taghead#1{\gdef\t@ghead{#1}\global\advance\t@ghd by 1}

\ifx\eqn\undefined
  \order{eqn}\let\@qn=\eqn
\else
  \let\eqn@=\eqn\order{eqn}\let\@qn=\eqn\let\eqn=\eqn@
\fi\let\ref@qn=\refeqn

\let\eqno@=\eqno
\def\eqno(#1){\eqno@({\rm\@qn{#1}})}

\ifx\eq\undefined
  \let\eq@=\relax
\else
  \let\eq@=\eq
\fi
\def\eq(#1){\eq@({\rm\@qn{#1}})}

\def\refeq#1{{\rm(\ref@qn{#1})}}

\newcount\r@fcount \r@fcount=0
\def\refcite#1{\each@rg\r@featspace{#1}}
\def\r@featspace#1#2 #3{\r@fcite#1#2,}
\def\r@fcite#1,%
  {\ifunc@lled{r@f}{#1}\global\advance\r@fcount by 1%
    \expandafter\xdef\csname r@fnum#1\endcsname{\number\r@fcount}%
    \expandafter\gdef\csname r@ftext\number\r@fcount\endcsname%
     {\message{ Reference #1 to be supplied }
      \if@iopp\refjl{}{}{}{Reference #1 to be supplied} \par
      \else Reference #1 to be supplied\par\fi}%
   \fi\csname r@fnum#1\endcsname}

\def\refis#1 #2\par
  {\ifunc@lled{r@f}{#1}
  \else
    \global\let\refjl=\numrefjl\global\let\refbk=\numrefbk
    \expandafter\gdef\csname r@ftext\csname r@fnum#1\endcsname\endcsname%
    {\writenewr@f#1>>#2\par}
  \fi}

\def\writenewr@f#1>>{}
\gdef\referencefile{\expandafter\immediate\csname newwrite\endcsname\reffile
                    \immediate\openout\reffile=\jobname.ref
   \def\writenewr@f##1>>%
  {\immediate\write\reffile{\noexpand\refis{##1}
     \expandafter\expandafter\expandafter\strip@t\expandafter
     \meaning\csname r@ftext\csname r@fnum##1\endcsname\endcsname}}
  \def\strip@t##1>>{}}

\newcount\r@fcurr
\def\listreferences{\global\r@fcurr=0
  {\loop\ifnum\r@fcurr<\r@fcount\global\advance\r@fcurr by 1   
   \if@iopp\let\refjl=\numr@@fjl\let\refbk=\numr@@fbk
     \csname r@ftext\number\r@fcurr\endcsname
   \else
     \numr@f{\number\r@fcurr}{\csname r@ftext\number\r@fcurr\endcsname}
   \fi\repeat}}

\newif\if@iopp
\ifx\numrefjl\undefined
  \@ioppfalse
\else
  \ifx\numrefbk\undefined
    \@ioppfalse
  \else
    \@iopptrue
  \fi
\fi

\if@iopp
  \message{ IOPP reference macros in use }
  \let\r@fjl=\refjl
  \let\numr@fjl=\numrefjl
  \let\r@fbk=\refbk
  \let\numr@fbk=\numrefbk
  \def\numr@@fjl#1#2#3#4{\numr@fjl{\number\r@fcurr}{#1}{#2}{#3}{#4}
     \let\refjl=\r@fjl\let\refbk=\r@fbk}
  \def\numr@@fbk#1#2#3{\numr@fbk{\number\r@fcurr}{#1}{#2}{#3}
     \let\refjl=\r@fjl\let\refbk=\r@fbk}
\else
  \message{ User defined reference macros }
  \def\numr@f#1#2 {\parindent=30pt
  \hang\noindent\rm {\hbox to 30truept{[#1]\hss\quad}}#2
  \par\parindent=16pt}
\fi

\catcode`@=12

\order{thm}\order{lem}


\overfullrule=0pt 

\def\e{{\rm e}}

\title{Exact solution of a one-dimensional fermion\break
 model with interchain tunneling}
\author{R. Z. Bariev$^{\rm (a)}$, A. Kl\"umper$^{\rm (b)}$, 
A. Schadschneider$^{\rm (c)}$
 and J.Zittartz$^{\rm (b)}$}
\address{Institut f\"ur Theoretische Physik, Universit\"at zu K\"oln,
Z\"ulpicher Strasse 77, D-50937 K\"oln, Federal Republic of 
Germany}
\vfill

\date

\beginabstract
A fermion model consisting of two chains with interchain tunneling is
formulated and solved exactly by the Bethe ansatz method. The interchain
tunneling leads to  Cooper pair like bound states and a threshold
energy is required to overcome the binding energy. The
correlation functions manifest superconducting properties.
\endabstract
Since several years there has been considerable interest in unusual mechanisms
for the formation of Cooper pairs because of the possible relevance 
for high $T_c$ superconductors. A simple model showing a pairing
transition was recently proposed in [1] to explain the properties of
the high-temperature cuprate superconductors. The underlying mechanism 
leading to the superconducting transition is an interlayer tunneling phenomenon.
Although the model in [1] is two-dimensional, it
is quite interesting to investigate its one-dimensional variant, as it 
has been conjectured [2] that properties of the one- and two-dimensional 
models have common aspects. Exact results in one dimension are 
often more easily available than in two dimensions and may provide a testing 
ground for approaches intended for more complex problems.

In this paper we present the exact solution to a new integrable model 
consisting of two chains with an interchain tunneling which may be 
considered as the simplest one-dimensional variant of the model in [1]. It 
will be 
shown that the attractive interaction leads to bound states of Cooper 
type pairs and the model has superconducting properties.
    
The Hamiltonian under consideration is the following
$$
{\cal H}=-\sum_{j=1}^L{\cal P}\!\!\left[c_{j1}^+c_{j+1,1}+
c_{j2}^+c_{j+1,2}-Vc_{j1}^+c_{j+1,1}^+\left(c_{j2}c_{j+1,2}+
c_{j-1,2}c_{j2}\right)+h.c.\right]{\cal P},\!\!\eqno(1)
$$
where the Fermi operator $c_{j\alpha}^+$ creates an electron at site $j$ with 
a sublattice index $\alpha (\alpha =1,2)$, $L$ is the number of lattice sites,
and $h.c.$ denotes the Hermitian conjugate. We impose the restriction of no 
simultaneous occupancies of sites on different sublattices which are nearest 
neighbours, i.e. $(j,1);\ (j,2)$ and $ (j,1);\ (j-1,2)$, and ${\cal P}$ is the 
projector on the subspace of allowed states. The Hamiltonian (1) 
contains two different terms. The first describes the hopping of 
individual electrons with transition rate set equal to 1, and the second 
describes the hopping of pairs 
of electrons with rate $V$ ($>0$). The competition of 
these terms is the main property of the model under consideration. 
It should be noted that this is similar to the one-dimensional model proposed 
by Penson and Kolb [3]. We hope that the study of (1) will also help 
to understand this nonintegrable model.

The exact solution for the eigenstates and eigenvalues of 
Hamiltonian (1) can be obtained within the framework of the Bethe 
ansatz method [4-6]. The structure of the Bethe ansatz equations follows from 
the solution of the two-particle problem. The nonvanishing elements 
of the $S$-matrix are 
$$\eqalign{
&S_{11}^{11}(k)=S_{22}^{22}(k)=i{\sinh\eta\over\sin(k+i\eta)} \e^{-ik},\cr
&S_{11}^{22}(k)=S_{22}^{11}(k)=-{\sin k\over\sin(k+i\eta)}\e^{-ik},\cr
&S_{12}^{21}(k)=S_{21}^{12}(k)=\e^{-ik},\quad \eta=-\ln V.\cr}
\eqno(2)
$$
A necessary and sufficient condition for the applicability of the Bethe
ansatz method are the Yang-Baxter equations [4]. Since 
the scattering matrix (2) satisfies these equations, our model (1) is 
integrable.

The discrete Bethe ansatz equations are derived following the standard 
procedure [7] by imposing periodic boundary conditions. Each state 
of the Hamiltonian is specified by one set of charge rapidities 
{$k_{j}$} representing the momenta of the electrons and one set 
of additional rapidities {$\Lambda_\alpha$}. All rapidities 
within a given set have to be different, corresponding to Fermi 
statistics. These rapidities are determined by the Bethe ansatz equations
$$\eqalign{
&\e^{ik_j L}\!=\!\prod_{l=1}^{N}\exp\left[{i\over2}
\left(k_{j}-k_{l}\right)\right]\!\prod_{\alpha=1}^{M}
{\sin[{1\over2}(k_j-\Lambda_\alpha)  
+{\i\over 2}\eta]\over\sin[{1\over2}(k_j-\Lambda_\alpha)  
-{\i\over 2}\eta]},\cr
&\prod_{j=1}^{N}{\sin[{1\over2}(\Lambda_\alpha-k_j)+
{\i\over 2}\eta]\over
\sin[{1\over2}(\Lambda_\alpha-k_j)-{\i\over 2}\eta]}=
-\prod_{\beta=1}^{M}{\sin[{1\over2}(\Lambda_\alpha 
-\Lambda_{\beta})+\i\eta]\over\sin[{1\over2}(\Lambda_\alpha -
\Lambda_{\beta})-\i\eta]}.\cr }
\eqno(3)
$$
The energy and the momentum of the corresponding state are given
by
$$
E=-2\sum_{j=1}^{N}\cos k_j\!,\quad 
P=\sum_{j=1}^{N}k_j.
\eqno(4)
$$

We remark that in contrast to [4-6] in our 
model the electron number on each separate sublattice is not 
conserved. In the case under consideration the
conserved quantities are the total number of electrons $N$ and the number 
$M$ of pairs of consecutive electrons which are on the same sublattice.

Each eigenstate of the Hamiltonian is specified by sets of 
rapidities which have to satisfy the Bethe ansatz equations (3). The 
structure of 
the solutions of these equations is different for $\eta<0$ and 
$\eta>0$. In the case $\eta>0$
all $k_{j}$ are real and particles move independently. There are no Cooper 
pairs in the system. Equations (3) are reduced to one set of 
equations for $k_{j}$
$$\eqalign{
&k_{j}L-{1\over2}\sum_{i=1}^{N}(k_{j}-k_{i})+
\sum_{i=1}^{N}\Phi(k_{j}-k_{i}) = 2\pi I_{j},\cr
&\Phi(k) = {k\over2} + \sum_{\nu=1}^{\infty}
{\exp(-\nu\eta)\over \nu} {\sin(\nu k)\over\cosh(\nu\eta)}.\cr}
\eqno(5)
$$
where $I_{j}$ are integer (half-integer) numbers for odd (even) $N$. 
(We have included the linear term $k/2$ in the definition of the function 
$\Phi(k)$ such that it is identical to the phase function used in [10].)
In the thermodynamic limit eqs. (4,5) are replaced by the integral equation 
for the distribution function of particles $\rho(k)$
$$
2\pi\rho(k) - \int_{-k_0}^{k_0}\varphi(k-k')\rho(k')d k' = 1-{1\over2}\rho,
\qquad \varphi(k)=\Phi'(k)
\eqno(6)
$$
with subsidiary condition for the particle density $\rho$
$$
\int_{-k_0}^{k_0} \rho(k)d k = \rho = N/L.
 \eqno(7)
$$

In the opposite case $\eta<0\ (V>1)$ pairs of electrons are energetically 
more favourable as compared to unpaired particles. In the ground state 
we only have Cooper pair like bound states with finite pairing 
energy which are characterized by pairs of complex wavenumbers
$$
k_{\alpha}^{\pm}= \Lambda_{\alpha}\pm\i\vert\eta\vert .
\eqno(9)
$$
Inserting (9) into the Bethe ansatz equations (3) we obtain the following 
set of equations
$$\eqalign{
&2L\Lambda_{\alpha} -2\sum_{\beta=1}^{M} (\Lambda_{\alpha}-\Lambda_{\beta})-
 \sum_{\beta=1}^{M} \Theta(\Lambda_{\alpha}-\Lambda_{\beta};\vert\eta\vert)=
2\pi J_{\alpha},\cr
&E=\sum_{\alpha=1}^{M}e_{0}(\Lambda_{\alpha}),\quad
e_{0}(\Lambda)= -4\cosh\eta\cos\Lambda,\cr
&\Theta(\Lambda;\eta)=2\arctan(\coth\eta\tan \Lambda/2),\quad
-\pi<\Theta(\Lambda ;\eta)\le\pi,\cr}
\eqno(10)
$$
where $J_{\alpha}$ are integers or half-odd integers depending on the parity
of $M$. In the thermodynamic limit eqs. (10) lead to the integral equation 
for the distribution function of Cooper pairs $\sigma(\Lambda)$:
$$
2\pi \sigma(\Lambda)+\int_{-\Lambda_0}^{\Lambda_0} \Theta'(\Lambda-\Lambda';
\vert\eta\vert)\sigma(\Lambda') d \Lambda' =2\left(1-{1\over2}\rho\right),
\eqno (11)
$$
with subsidiary condition
$$
\int_{-\Lambda_0}^{\Lambda_0} \sigma(\Lambda) d \Lambda ={1\over2}\rho.
\eqno(12)
$$

In order to understand the superconducting properties of the model, we 
shall investigate the long-distance behaviour of the correlation functions. 
For this purpose we shall use the methods of 
two-dimensional conformal field theory [8,9]. 
Our model has just one type of gapless excitations which correspond
to ``holes" and ``particles" in the $I$-distribution or $J$-distribution 
for the case $\eta >0$ and $ \eta <0$, respectively. 
Therefore, an application of this theory presents no difficulties and can be 
carried through as in [10,11]. The results of these calculations are the 
following.

For $\eta>0$ the long-distance power-law behaviour of the density-density 
correlation functions is given by
$$
\langle n(r)n(0)\rangle\simeq
\rho^2+Ar^{-\alpha} \cos(\pi\rho r),\quad
n(r)=\sum_{\tau =1,2}c_{r \tau}^+c_{r \tau}.
\eqno(13)
$$
The critical exponent $\alpha$ is expressed in terms of the dressed charge 
$Z$ which in our case is given by
$$
\alpha=Z^{2}/2,\quad Z=2\pi\rho(k_{0}).
\eqno(14)
$$
The superconducting properties of the system manifest themselves
in the behaviour of the 
pair correlation functions for which we obtain
$$
G_{p}(r,\tau) = \langle
c_{r1}^+c_{r+1,1}^+c_{0\tau}c_{1\tau}\rangle\simeq Br^{-\beta},\quad  
\beta=2/Z^2 ={1\over\alpha}.
\eqno(15)
$$

In the same way we can consider the case $\eta<0$. The expressions (13)-(15)
also hold in this case, but the dressed charge $Z$ is given in terms of the 
distribution function $\sigma(\Lambda)$
$$
Z=2\pi \sigma(\Lambda_{0}).
\eqno(16)
$$

Comparison of eqs.\ (13)-(16) with the corresponding 
results for the model of electrons
with correlated hopping [11] permits us to obtain a simple relation between 
the critical exponents of these models
$$\eqalign{
&\beta(\rho,\eta)=\left(1-{1\over2}\rho\right)^{-2}
\beta_{(c)}\left({1-\rho\over 1-{\rho/ 2}},\eta\right),\quad \eta>0, \cr
&\beta(\rho,\eta)=\left(1-{1\over2}\rho\right)^{-2}
\beta_{(c)}\left({\rho\over 2-\rho},\vert\eta\vert\right),\quad \eta<0, \cr
&0<\rho<1,\cr}
\eqno(17)
$$
where $\beta_{(c)}$ is the critical exponent of the superconducting 
correlation function for electrons with correlated hopping [11]. Numerical
results for the critical exponent $\beta$ are shown in Fig. 1.

In addition to (17) there is a relation
of the charge stiffnesses $D(\rho,\eta)$ and $D_{(c)}(\rho,\eta)$ 
of both models
$$\eqalign{
&D(\rho,\eta)=\left(1-{1\over2}\rho\right)\e^{-\eta}
D_{(c)}\left({1-\rho\over 1-{\rho/ 2}},\eta\right),\quad \eta>0, \cr
&D(\rho,\eta)=\left(1-{1\over2}\rho\right)
D_{(c)}\left({\rho\over 2-\rho},\vert\eta\vert\right),\quad \eta<0, \cr
&0<\rho<1.\cr}
\eqno(18)
$$
Figs. 2 and 3 show numerical results for the charge stiffness 
and transport mass of the charge carriers defined by
$${m\over m_e}={D^0\over D},\eqno(19)$$
where $m_e$ is the bare mass and $D^0={2\over\pi}\sin{\pi\over 2}\rho$ 
is the charge stiffness of free fermions. For densities $\rho$ approaching
the maximum allowed value 1 the masses diverge due to a (trivial) 
metal-insulator transition.

The existence of a simple relation between the critical exponents,
$\alpha\beta=1$, is certainly a consequence of the universal Luttinger 
behaviour of critical models with one type of gapless excitations [12]. 
As our model of interchain tunneling (1) is described by a conformal field
theory with central charge $c=1$, the critical exponents depend in a 
nonuniversal way on the particle density $\rho$ and the interaction strength
$\eta$. Although the model of electrons with correlated hopping [11] is
different, there exists a mapping of the phase diagram of the first model to
the latter one with a simple relation of the phase points and the corresponding
values of the exponents $\beta$ and $\beta_{(c)}$, respectively. 
Note that the mapping is one-to-one, however, the physical properties of the
models differ.
In contrast to [11] we see that for (1) with couplings $\eta>0$ there
are no densities $\rho$ for which $\beta<\alpha$.
This means that the dynamics is dominated by the motion of individual 
electrons rather than by the motion of pairs, and we have 
no ``superconducting" phase for any particle concentration.

For $\eta <0$ the situation is quite different. As in the 
model with correlated hopping there is a regime for which 
$\beta <\alpha$, see. Fig. 4. In this case the pair correlation function 
has a slower decay than the density-density correlation function and 
thus dominates. In one-dimensional systems which do not 
undergo a condensation and do not have long-range order such behaviour of 
correlation functions is the closest analogy to the existence of 
a superconducting phase [13].

\ack{This work has been performed within the 
research program of the Sonderforschungsbereich 341, K\"oln-Aachen-J\"ulich.
RZB gratefully acknowledges the hospitality
of the Institut f\"ur Theoretische Physik, Universit\"at zu K\"oln.}

\references

\item{(a)}Permanent address: The Kazan 
Physico-Technical Institute of the Russian Academy of Sciences,
Kazan 420029, Russia

\item{(b)}Email: kluemper@thp.uni-koeln.de,
zitt@thp.uni-koeln.de

\item{(c)}Present address: Institute for Theoretical Physics, 
State University of New York at
Stony Brook, Stony Brook, N Y 11794-3840, U.S.A.

\item{[1]} S.Chakravarty, A.Sudbo, P.W.Anderson and S.Strong, Science 261, 337 (1993)

\item{[2]} P.W.Anderson, Phys.Rev.Lett. 64, 1839 (1990)

\item{[3]} K.Penson and M.Kolb, Phys.Rev.B 33, 1663(1986)

\item{[4]} C.N.Yang, Phys.Rev.Lett. 19, 1312 (1967)

\item{[5]} E.H.Wu and F.Y.Wu, Phys.Rev.Lett. 20, 1445 (1968)

\item{[6]} B.Sutherland, Phys.Rev.B 12, 3795 (1975)

\item{[7]} L.A.Takhtajan and L.D.Faddeev, Russ.Math.Surveys 34, 11(1979)

\item{[8]} A.A.Belavin, A.M.Polyakov, A.B.Zamolodchikov, Nucl.Phys.B 241,
333 (1984)

\item{[9]} J.L.Cardy, Nucl.Phys.B 270 [FS16], 186 (1986) 

\item{[10]} R.Z.Bariev, A.Kl\"umper, A.Schadschneider
and J.Zittartz,  J. Phys.A: Math.Gen. 26, 4863 (1993)

\item{[11]} R.Z.Bariev, A.Kl\"umper, A.Schadschneider
and J.Zittartz,  J. Phys.A: Math.Gen. 26, 1249 (1993)

\item{[12]} L.P.Kadanoff and A.Brown, Ann. Phys. 121, 318 (1979)

\item{[13]} N.M.Bogoliubov and V.E.Korepin, Int.J. Mod.Phys.A 3,
427 (1989)  

\figures

\figcaption{(a) Depiction of the critical exponent $\beta$ in dependence
on the particle density $\rho$ for different values of $\eta=$0.1,
0.5, 1, 10. (b) The same as for (a) with negative values of $\eta$, $|\eta|=$
0.05, 0.1, 0.5, 1, 10.}

\figcaption{Charge stiffness $D_c$ and effective mass $m$ of 
the charge carriers for different interaction strengths 
$\eta=$ 0.1, 0.5, 1, 2, 10.}

\figcaption{Charge stiffness $D_c$ and effective mass $m$ 
for negative values of $\eta$, $|\eta|=$ 0.1, 0.5, 1, 2.}
 
\figcaption{Depiction of the phase diagram for $\eta<0$. The 
solid line separates regions with dominating pair and dominating 
density-density correlations ($\beta<\alpha$ and $\beta>\alpha$), 
respectively. For large $\eta$ the critical density approaches 
$(4-2\sqrt{2})/3=0.39...)$.}
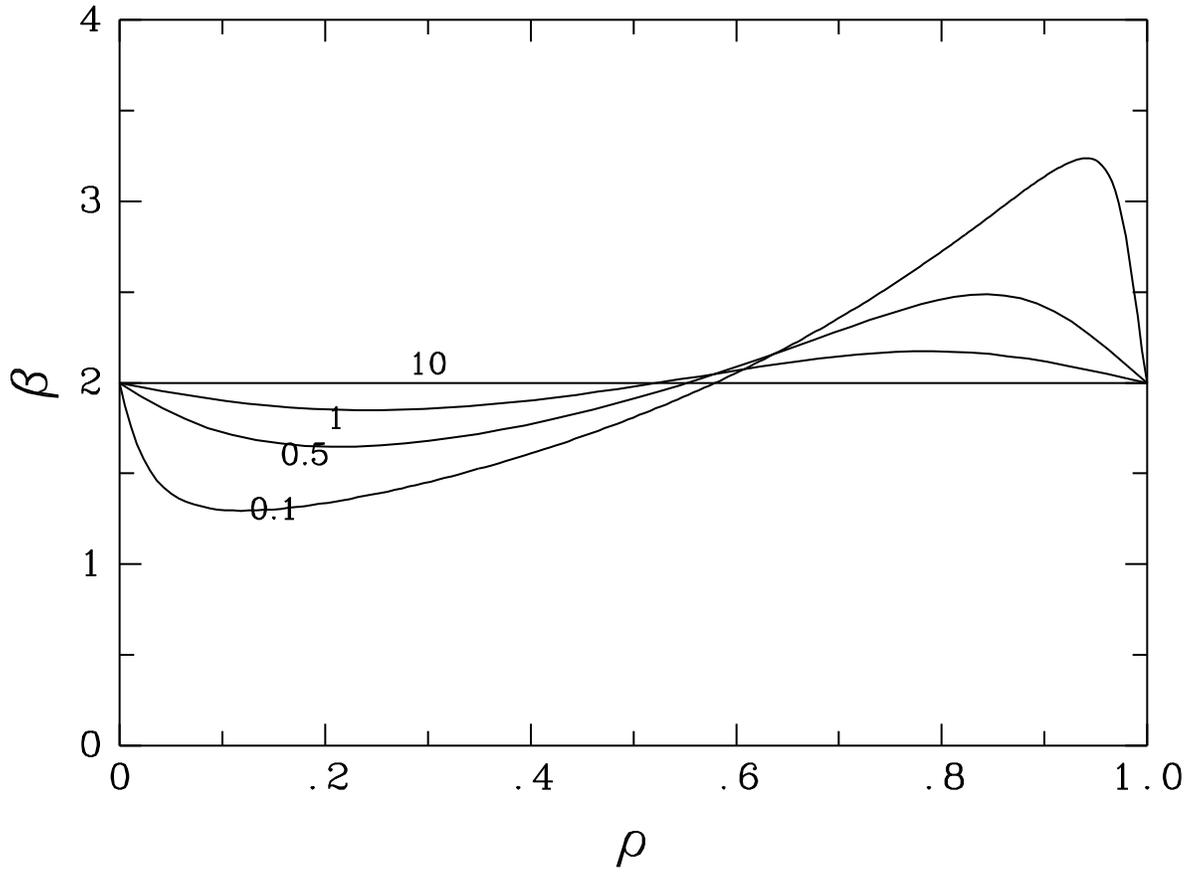
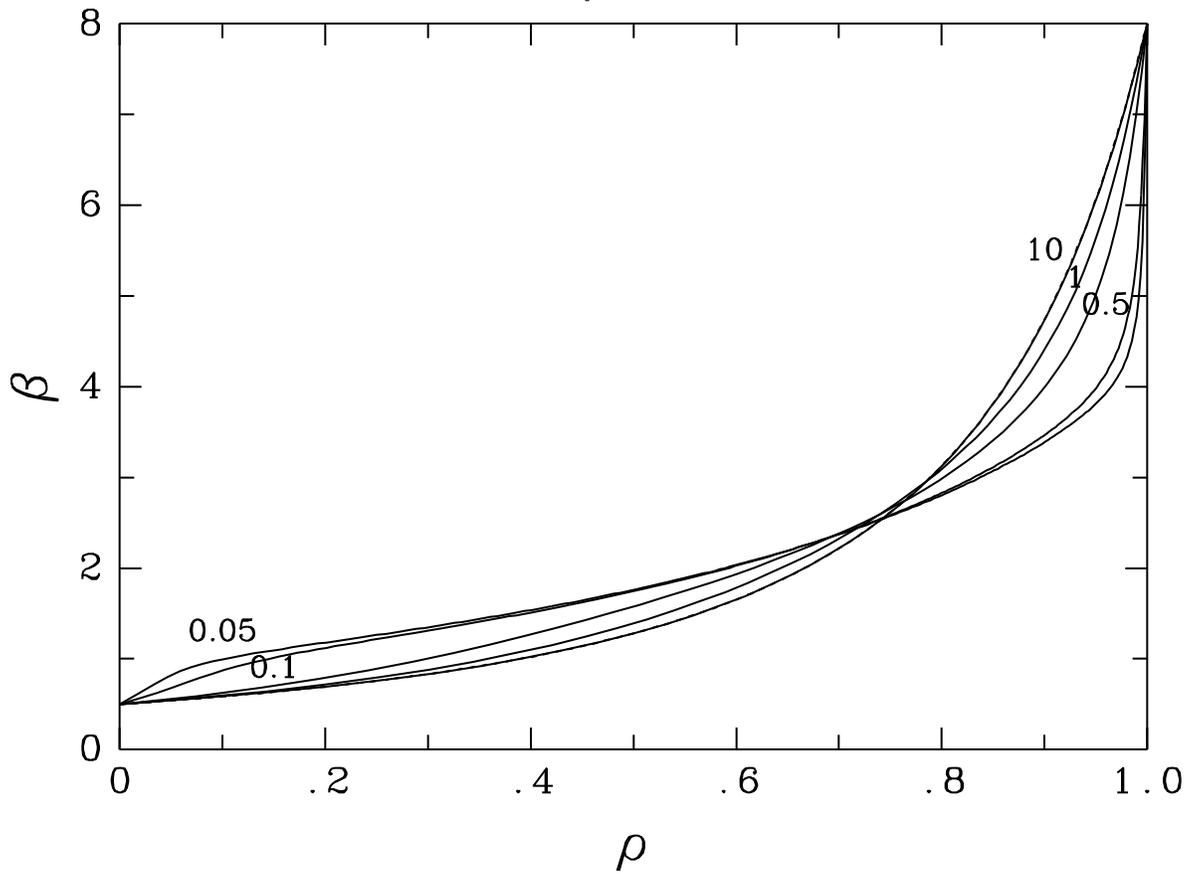
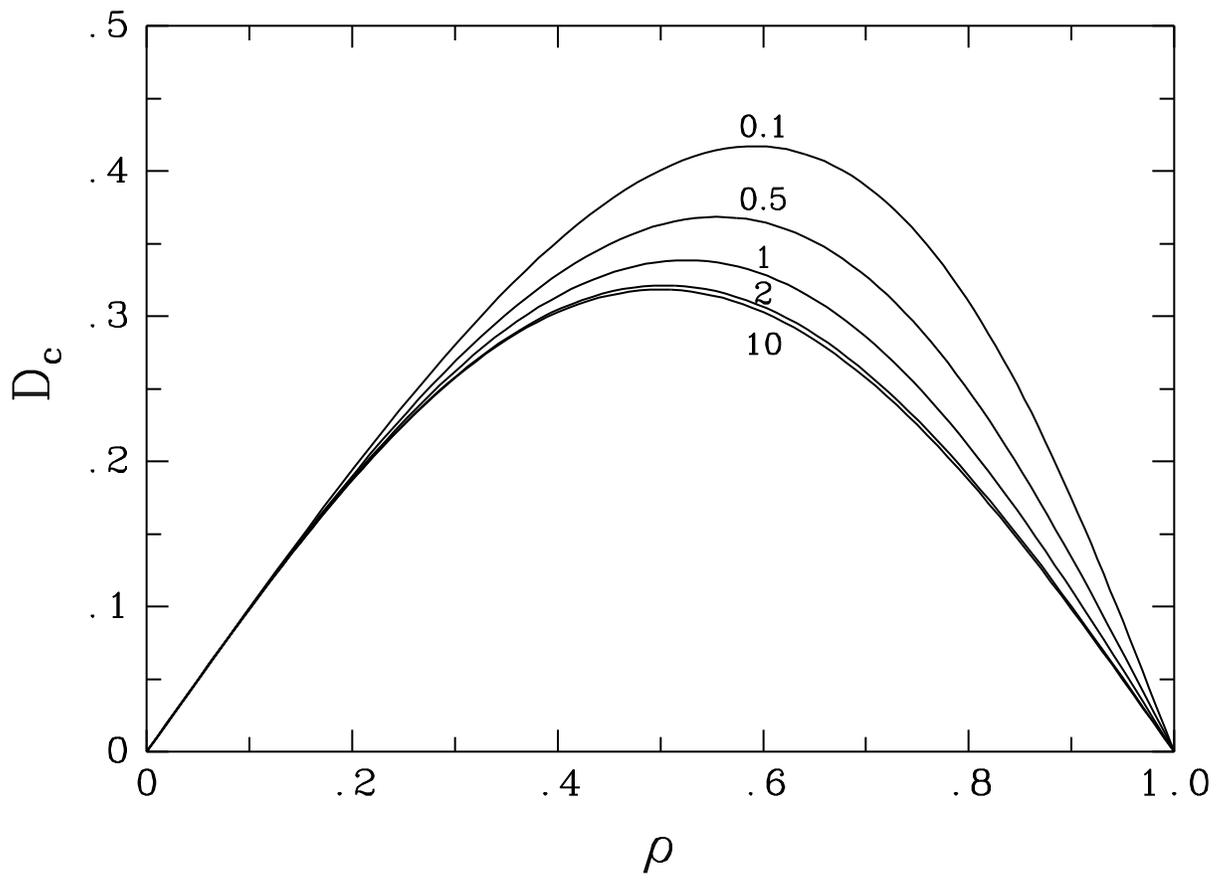
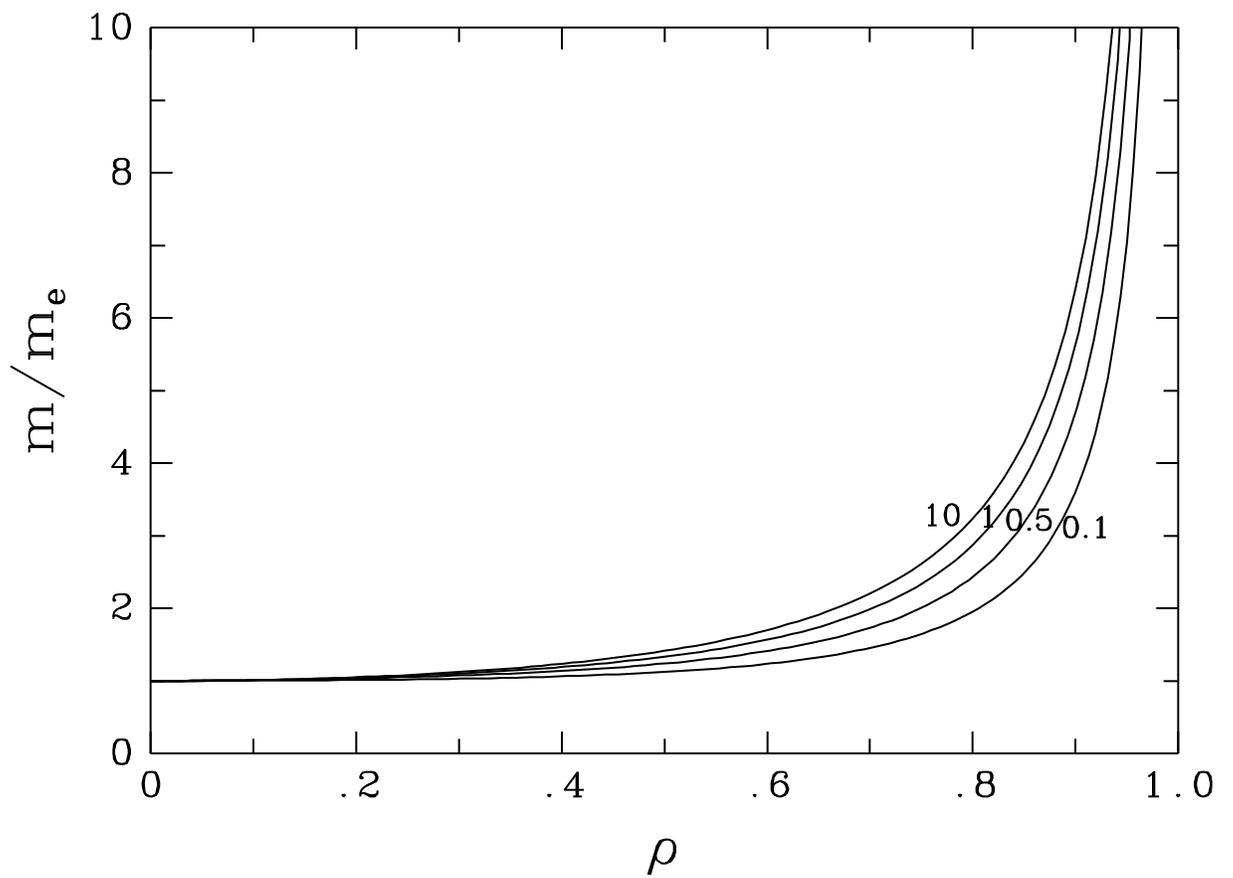
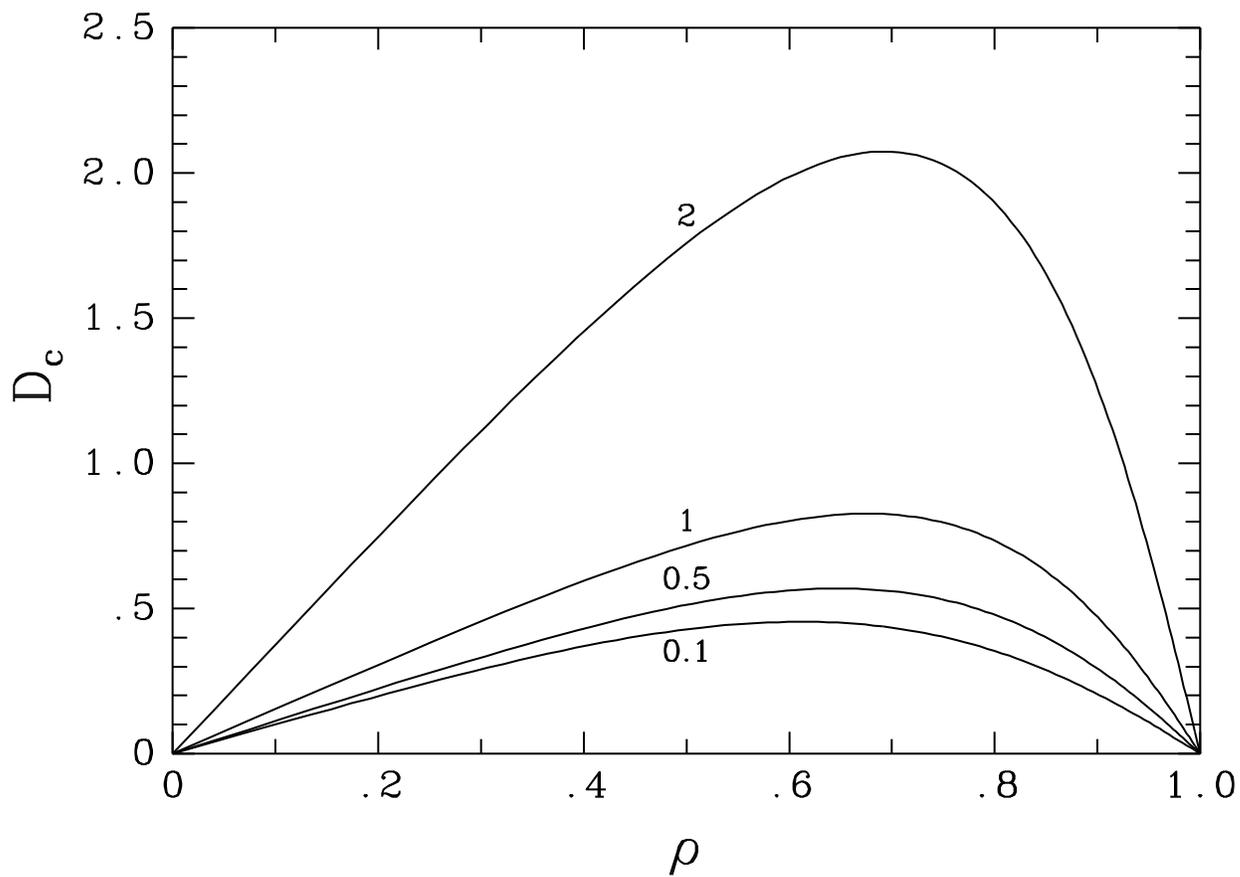
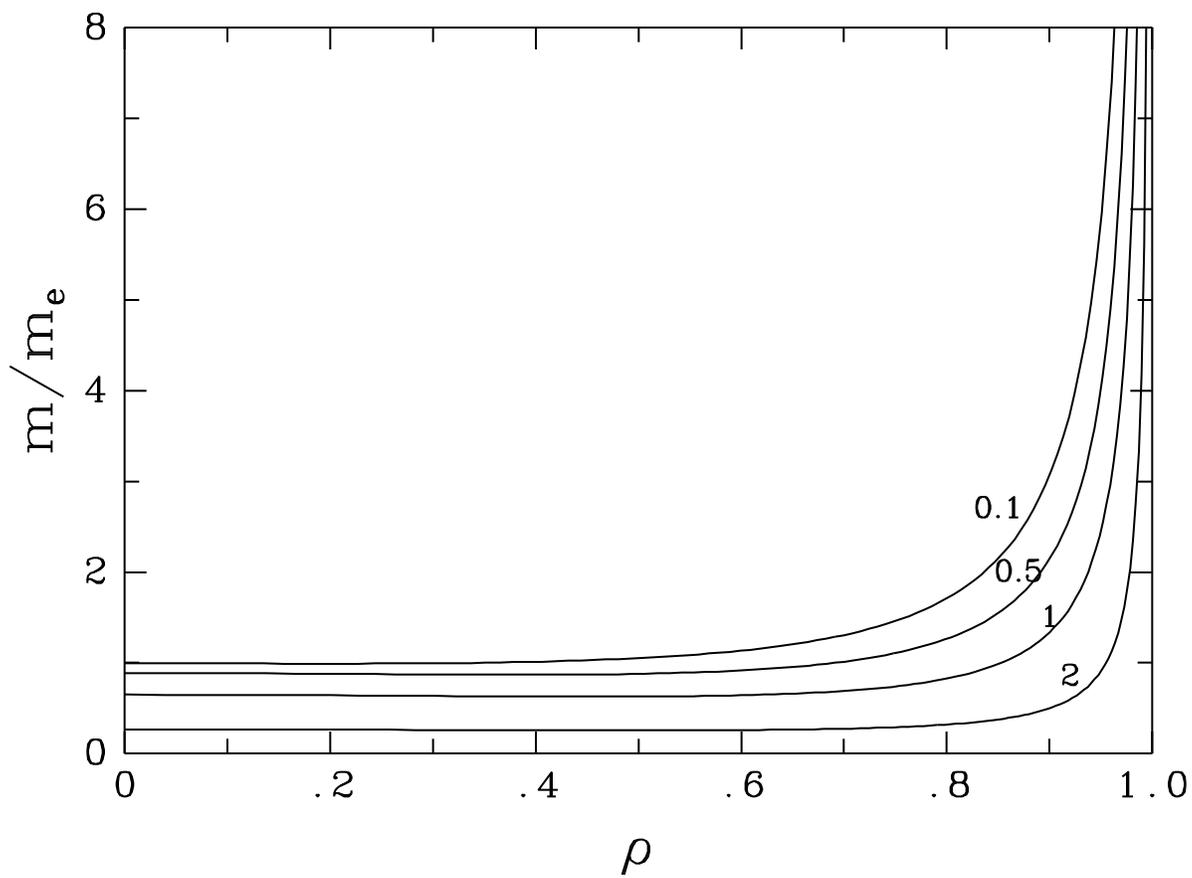
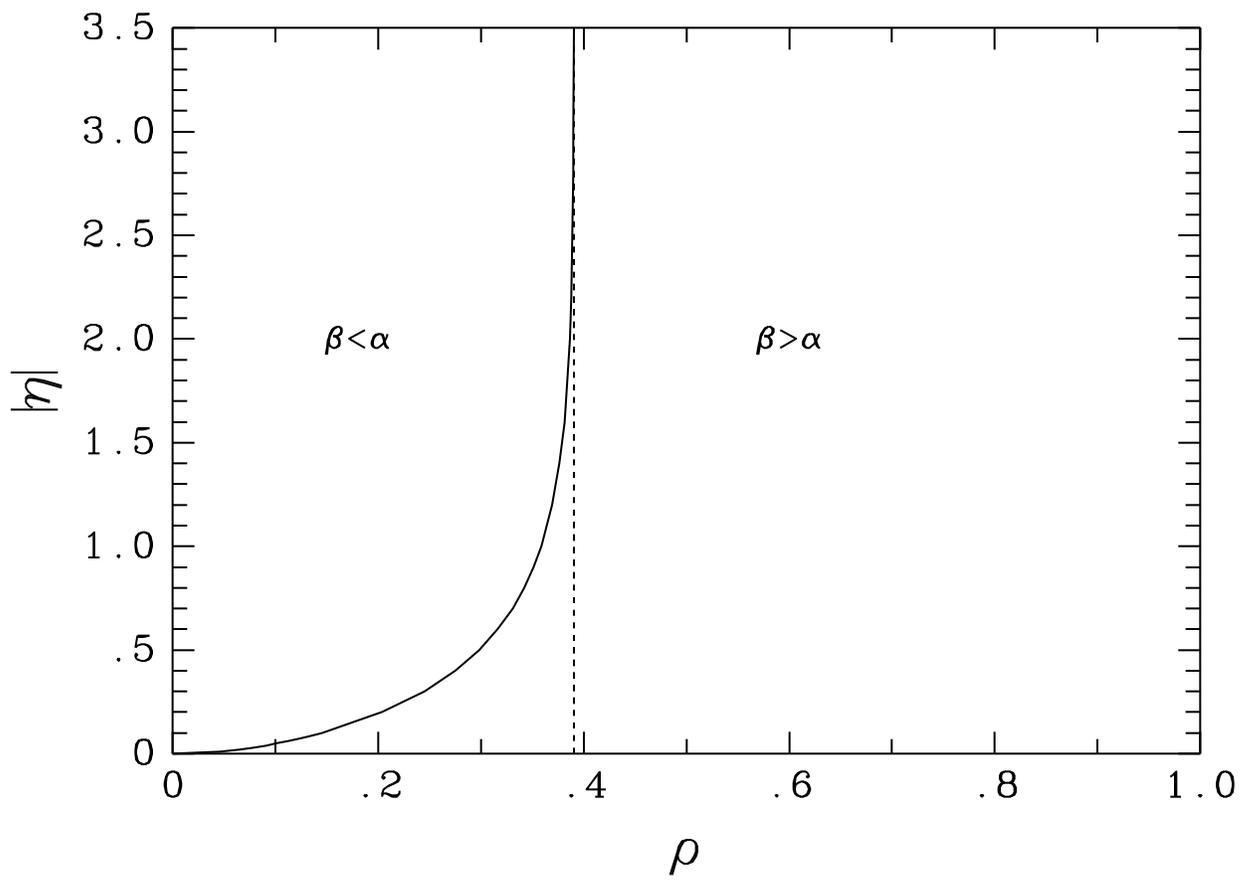
\bye